\documentclass[12pt]{article} 
\usepackage{amsmath,amssymb}
\usepackage{graphicx}
\usepackage{epsf,rotate}
\usepackage{epsfig}
\usepackage{rotating}
\begin{document}
\title{Critical behavior of a non-equilibrium
interacting particle system driven by an oscillatory field.}
\author{Roberto A. Monetti and Ezequiel V. Albano \\
Instituto de Investigaciones Fisicoqu\'{\i}micas Te\'oricas y 
Aplicadas (INIFTA) \\ UNLP, CONICET, CIC (Bs. As.) \\  C. C. 16 Suc. 4,
1900 La Plata, Argentina.}

\maketitle
\noindent Pacs 64.60.Cn \{Order-disorder transitions. Statistical Mechanics
of model systems.\} \\
Pacs 82.20.Mj \{Nonequilibrium kinetics.\} \\
Pacs 66.30.Hs \{Self-diffusion and ionic conduction in nometals.\}
\begin{abstract}
First- and second-order temperature driven transitions 
are studied, in a lattice gas driven by an oscillatory field.
The short time dynamics study provides
upper and lower bounds for the first-order transition points
obtained using standard simulations. The difference
between upper and lower bounds is a measure
for the strength of the first-order transition and
becomes negligible small for densities close to one
half. In addition, we give strong evidence on the 
existence of multicritical points and a critical temperature
gap, the latter induced by the anisotropy introduced 
by the driving field. 
\end{abstract}
\newpage
Far from equilibrium systems (FFES) are ubiquitous in nature and their
theoretical understanding will
contribute to the progress of scientific areas
in physics, chemistry, biology, ecology, economy, etc. 
Since the
theoretical development of non-equilibrium statistical
mechanics is still in its infancy, a useful approach to FFES
is to study simple models by means of various techniques such as 
numerical simulations, mean-field approximations, phenomenological
scaling, field-theoretical developments, etc. Within the
broad context of FFES,
driven diffusive systems (DDS) \cite{kls} have very recently
received growing attention 
\cite{ktl1,wan,joa3};
for reviews see e.g. \cite{zia2,zia3,joa5}.  
The classical model for DDS was proposed by Katz et al. (KLS) \cite{kls}
and is based on the equilibrium Ising model \cite{isi}. 
Using the lattice gas
language, the KLS model introduces an external 
driving field to the Ising model.
However, due to this modification,
the system now evolves towards a non-equilibrium stationary-state 
(NESS). In spite of considerable effort
devoted to the study of the KLS model, 
there are still controversies on the
understanding of numerical data \cite{ktl1,wan,joa3} and  
its theoretical description is the subject of
an ongoing debate \cite{grana,vir,grana1}. 

In this work we study a DDS subjected to the action of 
an oscillating driving
field. One of the motivations for this approach is that
the periodical field can be realized in numerous practical
applications such as charged colloids
between the plates of a capacitor \cite{con},
electrophoresis experiments in pulsed fields \cite{elec}, 
gas condensation in the presence of
ultrasonic waves \cite{soun}, segregation of granular materials
in vibrating containers, 
etc.

The aim of this work is to perform an extensive simulation study of 
the dependence of the temperature-driven transitions  of the
model on both the density of particles
and the magnitude of the field. Measurements of stationary
properties combined to an study of the short time dynamics
allow us to drawn a detailed phase diagram of the model 
that lead us to  the discovery of a 
multicritical point. Furthermore, we developed a coupled
mean field approach that yields results in agreement with
the simulations.

The model is defined on the 
square lattice assuming a rectangular
geometry $L_x, L_y$, using ``brick wall'' (periodic) 
boundary conditions across (along) the $y-$ ($x-$)axis where
the oscillatory field is applied, respectively.
A lattice configuration $\eta$ is specified by the set of
occupation numbers $n_{i,j} = \{0,1\}$, corresponding to each
site of coordinates $(i,j)$, i.e. $\eta = \{n_{i,j}\}$. 
Nearest-neighbor
attraction with a coupling constant $J > 0$, 
is considered. So, in the absence of 
a field the Hamiltonian $\cal H$ is given by
\begin{equation}
{\cal H} = -4J \sum_{<ij,i'j'>} n_{i,j} n_{i',j'}       , \label{ham}
\end{equation}
\noindent where the summation is over nearest-neighbor sites only.
The driving oscillatory field $E$ acts along the $\pm y-$direction 
with half-period $\tau$.
The coupling to a thermal bath at temperature $T$
and the action of the field are
considered through Metropolis jump rates,
namely $min [ 1, exp -( \{\Delta {\cal H} - \epsilon E(\tau)\} /k_{B}T )$,
where $k_{B}$ is the Boltzmann constant, 
$\Delta \cal H$ is the change in $\cal H$
after the exchange, and $ \epsilon = (-1,0,1)$ for
a particle attempting to hop (against, orthogonal, along)
the driving field, respectively.
For $E = 0$ and half-filled lattices,
the model reduces to the 
Ising model in absence of magnetic field. In the
thermodynamic limit the Ising model exhibits a second-order
phase transition at a temperature
$T_{c}^{I} = 2.2692..J/k_{B}$.

Monte Carlo simulations are performed on lattices of aspect ratios
$L_{x}/L_{y} = 2$ and $1$, with $30 \leq L_{y} \leq 480$. 
$T$ is reported in units of $J/k_{B}$ and $E$ is given
in units of $J$. The starting
configuration is obtained by randomly filling the sample
with probability $\rho_{o}$, which is also the density
of particles that remains constant.
One Monte Carlo time step (mcs) involves
$L_{x} L{y}$ trials. Data are obtained disregarding
$10^{6}$ mcs in order to allow the system to reach a NESS,
and averages are taken over the subsequent $10^{6}$ mcs.
Using this procedure a single data point, as e.g. shown
in figure 2, requires $\approx 1$ day of CPU time
in an AMD 700 MHz  processor. 

The model has also been studied by means of a  
coupled mean-field (CMF) approach.
In order to write down the CMF equations 
the local density of particles $\rho_{i,j}$ at site
$(i,j)$ is defined which is the probability of 
finding a particle in this site.
Due to normalization, one has $\rho_{i,j} + h_{i,j} = 1$, where
$h_{i,j}$ is the probability for the site $(i,j)$ to be empty.
Then, one has to consider all
events that may cause $\rho_{i,j}$ to change. 
$\rho_{i,j}$ may
increase by the arrival of particles due to unbiased (biased)
diffusion perpendicular (parallel) to the driving field, respectively.
Similarly, the density may decrease by an outgoing flux of particles
to neighboring sites. The implementation of the CMF
leads to a set of $L_{x} L{y}$ coupled non-linear 
differential equations. Here, we will only sketch out the form of such
equations for the sake of space. Let $\eta' [(i,j);(i', j')]$
be the configuration obtained from $\eta$ by interchanging the
content of site $(i,j)$ with that of a neighboring site $(i',j')$.
Then, the Metropolis rates  are functions $F$ of 
${\cal H}(\eta') - {\cal H}(\eta) - \epsilon E(\tau) = 
\Delta {\cal H}[(i,j);(i',j')] - \epsilon E(\tau) $.   
So, $\rho_{ij}$ evolves in 
time according to:
\begin{eqnarray}
\frac { d\rho_{i,j} }{dt} = 
h_{i,j} \{ \rho_{i+1,j} F\{ \Delta {\cal H}[(i,j);(i+1,j)],T \} 
+ \label{mf} 
\rho_{i-1,j} F\{ \Delta {\cal H}[(i,j);(i-1,j)],T \} + \\ \nonumber
\rho_{i,j+1} F\{ \Delta {\cal H}[(i,j);(i,j+1)],T,E(\tau) \} + 
\rho_{i,j-1} F\{ \Delta {\cal H}[(i,j);(i,j-1)],T,E(\tau) \} \} - \\ \nonumber
\rho_{i,j} \{ h_{i+1,j} F\{ \Delta {\cal H}[(i,j);(i+1,j)],T \} + 
h_{i-1,j} F\{ \Delta {\cal H}[(i,j);(i-1,j)],T \} +  \\ \nonumber
h_{i,j+1} F\{ \Delta {\cal H}[(i,j);(i,j+1)],T,E(\tau) \} + 
h_{i,j-1} F\{ \Delta {\cal H}[(i,j);(i,j-1)],T,E(\tau) \} \}. 
\end{eqnarray}
\noindent The set of equations (3) is solved numerically
starting from a random initial distribution of particles
and using an integration time step of $\Delta t = 0.25$,
in arbitrary units. Numerical integrations are performed until
$t = 25000$ and averages are taken for $t \geq 20000$.   
In the CMF approach the 
excluded volume interaction is taken into account
in a probabilistic way and stochastic fluctuations are disregarded, 
in contrast to the 
Monte Carlo method which has intrinsic fluctuations and 
excluded volume is deterministically satisfied. 
However, the CMF approach is derived directly form
the microscopics, so it contains the same symmetries
than the lattice gas model.
One advantage of the CMF method is that one can
obtain the spatial mass distribution. In fact,
figure 1 corresponds to a NESS where 
a multi stripped pattern is observed.
\begin{figure}
\centerline{\epsfig{file=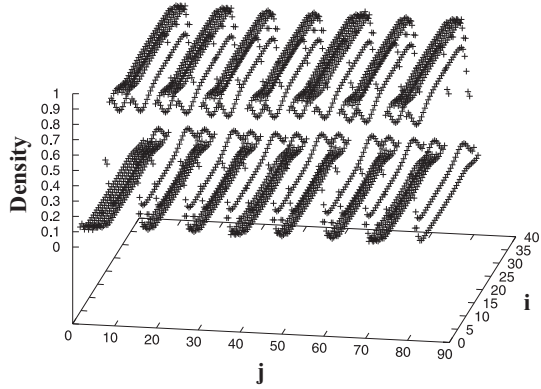, height=2.0in}}
\caption{3d plot of the density distribution 
characteristic of a NESS multi-stripped configuration
obtained with the 
CMF method. $L_{x} = 80$, $L_{y} = 40$, $T = 2.0$, 
$\rho_{o} = 0.50$, $E = 10$ and $\tau = 10$.}
\label{f.1}
\end{figure}
An intriguing feature of driven dissipative systems
is the occurrence of highly ordered
and complex patterns as shown in figure 1. 
Since the system constantly gains (loss) energy from (to) the external
field (thermal bath), respectively, the observed stationary
states are by no means equilibrium states. In fact, they are
truly {\bf non-equilibrium steady states}.

In order to perform a quantitative investigation, the
longitudinal order parameter ($OP_{x}$) is defined 
as the excess density, namely
\begin{equation}
OP_{x} \equiv (R L_{x})^{-1}
\sum_{i=1}^{L_{x}} |P(i) -\rho_{o}|   ,  \label{op1}
\end{equation}
\noindent where $P(i) = (L_{y})^{-1} \sum_{j=1}^{L_{y}} n_{ij}$
is the density profile along
the $x-$direction and $R = (2\rho_{o}(1 - \rho_{o}))$ is
a normalization constant.
Similarly, 
$OP_{y}$ can also be defined.

The dependence of the nature of the ordered phase 
on the period of the applied field has been 
investigated \cite{jsp}. For temperatures below criticality,
it is found that for short periods (say $\tau < 4 L_{y}$)
the system exhibits NESS with stripped patterns such as that
shown in figure 1. However, for larger periods  (say $\tau > 4 L_{y}$)
the system alternates between almost equilibrium states (AES)
such as those corresponding to molecules in a gravitational 
field. The crossover from NESS to AES has a characteristic time
of the order of  $\tau \approx 4 L_{y}$. In this work,
we are interested
in the critical behavior of NESS so we have restricted ourselves
to the case  $\tau = 10$ mcs, without loosing generality
because the same behavior will be valid for periods
such as $\tau < 4 L_{y}$ for finite lattices and all
periods in the thermodynamic limit. 
So, $\tau$ plays an important role in this model. In fact, for the case treated
in this work, namely $\tau < 4 L_{y}$, $OP_{x}$
is a well defined quantity independent of time $t$. However, for $\tau > 4
L_{y}$, $OP_{x}$ and $OP_{y}$ are functions of time $t$, since the
system alternates between AES as mentioned above. So, the 
half-period changes the nature of the problem and a crossover from NESS
to AES is observed \cite{jsp}.
In addition, since the oscillatory field causes the 
current of the driven gas averaged over long times to vanish, the
symmetries of the model are different from those of the KLS model.
From the theoretical point of view, this fact is essential
to establish the universality class of the model, as
will be discussed below.     

Figure 2(a) shows results obtained for $E = 1$.
For low densities ($\rho_{o} \leq 0.15$) the observed transitions
are abrupt and exhibit strong metastability, so they
are first-order. In contrast, for $\rho_{o} \geq 0.40$ one
observes second-order or very weak first-order like  behavior. 
Notice that for $\rho_{o}= 0.20$ and $\rho_{o}= 0.40$ we have
also included data which demonstrate the particle-hole
exchange-invariance of the results. The existence of both
first- and second-order transitions can also be observed by 
using the CMF approach. These results are in excellent 
agreement with Monte Carlo data, as shown in the inset of figure 2(a).
Figure 2(b) and 2(c) show that for low densities,
$T_c$ depends on the 
amplitude of the field, so that the higher the field
the lower the $T_c$. Furthermore, these figures also reveal that 
weaker first-order transitions are obtained for
smaller amplitudes of the field. Remarkably, results obtained by means
of the CMF approach exhibit the same decreasing trend than the Monte
Carlo data. 
\begin{figure}
\epsfig{height=2.5in, width=2.0in, file=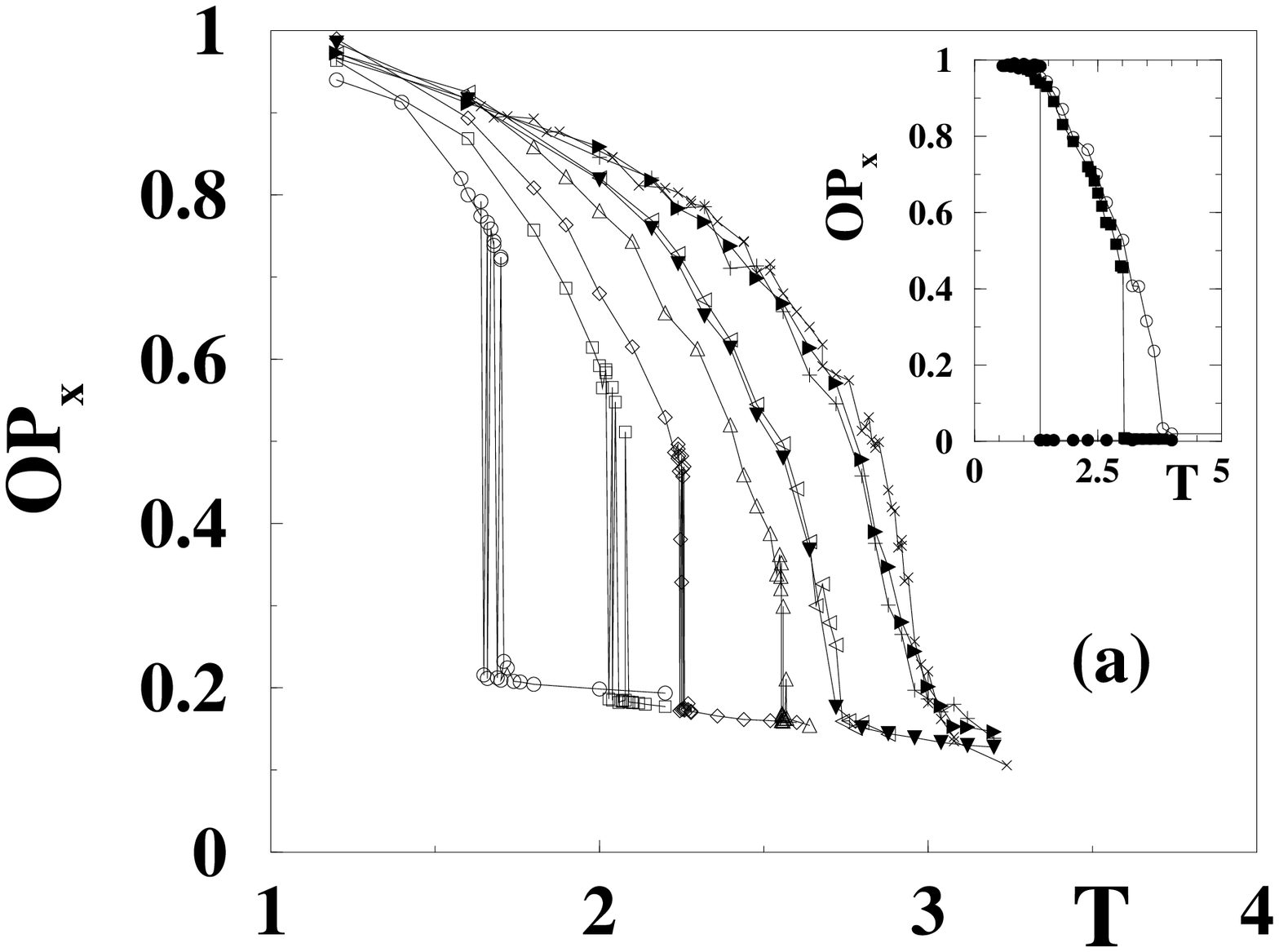}
\epsfig{file=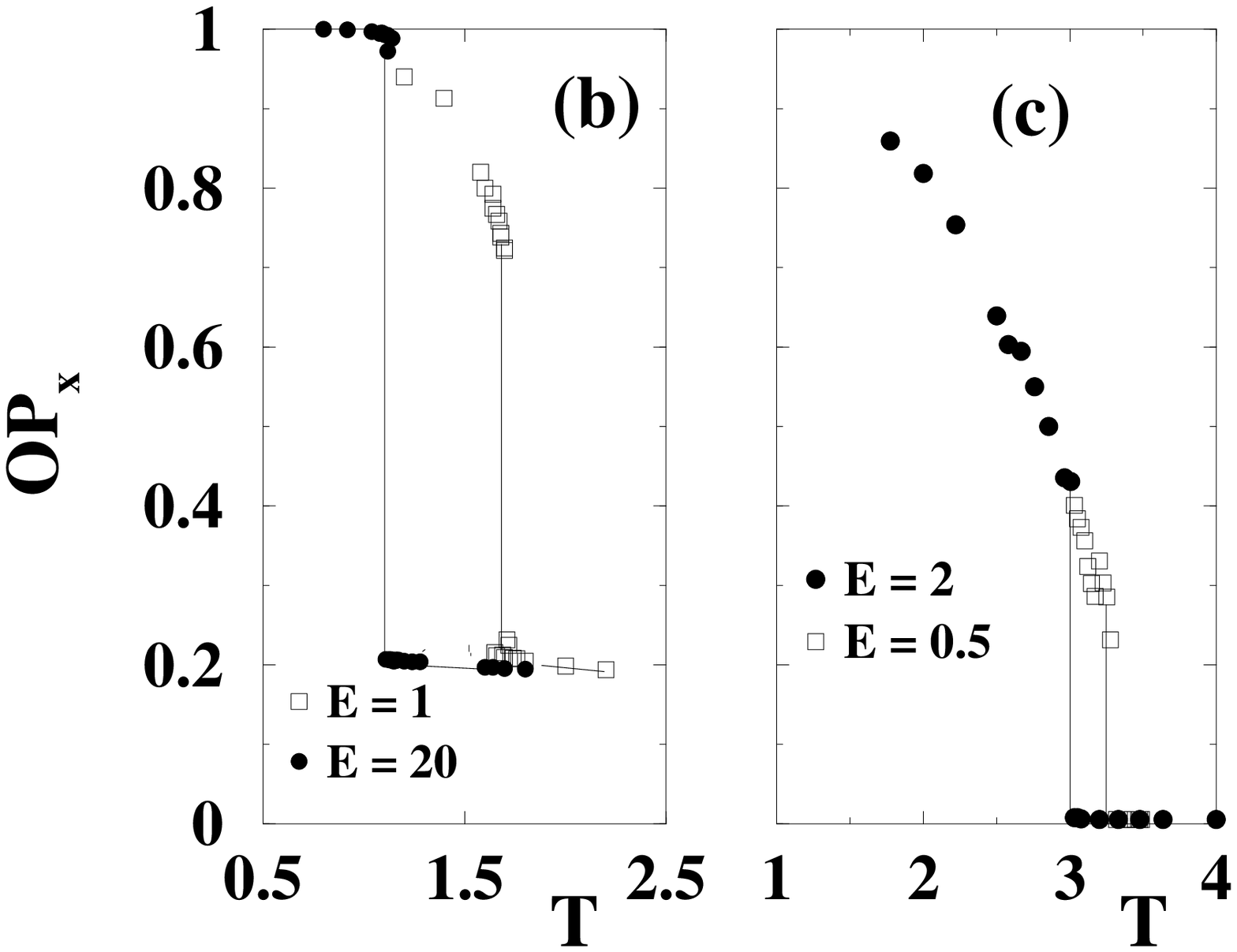, height=2.5in}
\caption{(a) Plots of $OP_x$ versus $T$ obtained for 
$L_{x} = 240$, $L_{y} = 120$, 
$E = 1$, $\tau = 10$ mcs and different values of 
$\rho_{o}$: $\circ, \rho_{o} = 0.05$;
$\square, \rho_{o} = 0.075$; $\bigtriangleup, \rho_{o} = 0.10$;
$\bigtriangledown, \rho_{o} = 0.15$; $+, \rho_{o} = 0.20$; 
$\blacksquare, \rho_{o} = 0.80$; $\blacktriangle, \rho_{o} = 0.40$; 
$\bullet, \rho_{o} = 0.60 $ and $\star, \rho_{o} = 0.50$. 
The inset shows results obtained solving the CMF equations
for $E = 10$ and $\tau = 1$. 
$\bullet$ $\rho_{o} = 0.15$, 
$\blacksquare, \rho_{o} = 0.30$
$\circ$ $\rho_{o} = 0.50$.
(b) Plots of $OP_x$ versus $T$ obtained by using the 
Monte Carlo method for $\rho_{o} = 0.05$ and 
the values of the field indicated in the figure.
(c) As in (b) but solving the CMF equations for $\rho_{o}=0.30$.}
\label{f.2}
\end{figure} 

Figure 3 shows the phase diagrams obtained for a fixed lattice size
($L_{x}/L_{y} = 2, L_{y} = 120$) and two values of the driving field,
namely $E = 1$ and $E = 50 \approx \infty $.  
\begin{figure}
\centerline{\epsfig{file=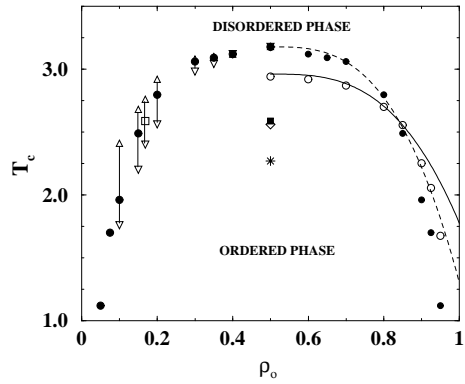, height=2.0in}}
\caption{Phase diagram, $T_{c}$ vs $\rho_{o}$, obtained
from the data of figure 2(a). Empty (filled) symbols 
correspond to $E = 1$ ($E = \infty$), respectively.
On the left side, the symbols $\bigtriangleup$ and 
$\bigtriangledown$ show the upper and lower bounds
for $T_{c}$ as obtained by means of the short time dynamics study
for $E = \infty$. $\square$ shows the location of the multicritical point.
The full and the dashed  curves, drawn on the right side, correspond
to the best fit of the data obtained using eq. (\ref{eq9}).
$\ast$, $\blacksquare$ and $\diamond$ show the location of the
critical temperature of the Ising model
$T_{c}^{I}$ , the critical temperature predicted 
by eq. (\ref{eq9}) for $E \rightarrow 0$, and 
the lower bound obtained for $E = 0.01$ using the short time dynamic
analysis, respectively.}
\label{f.3}
\end{figure}
Using a method recently proposed for the study of the
short time dynamics of weak first-order transitions \cite{boz}
it is possible to determine both lower and upper bounds for $T_{c}(\rho_{o})$
valid in the thermodynamic limit and further generalize the phase 
diagram for $E > 1$.
The idea behind the proposed method
is based on the existence of two pseudo critical points at
$T^{*}$ and $T^{**}$ near the weak first-order transition
point $T_{c}$ with  $T^{*} < T_{c} < T^{**}$. These points can
be obtained accurately from two short time dynamical processes
starting from fully disordered and zero temperature states, respectively.
In second-order transitions $T^{*}$ and $T^{**}$ overlap with the
transition point $T_{c}$, so the difference between
$T^{*}$ and $T^{**}$ also gives a criterion for the 
weakness of the first-order transition \cite{boz}.
Consider a system at $T < T_{c}(\rho_o)$ and the evolution process
from a fully disordered state. Due to the geometrical constrained
$L_{x}/L_{y}^{\phi} \gg 1$ ($\phi \approx 0.2$) \cite{muka} 
configurations at short times
exhibit multi-stripped patterns that are long lived, only relaxing to
the single stripe state after a time of the order $t \sim L_{x}^{3}
L_{y}$ \cite{muka}. Even in the case of square geometry, both
the present model and the KLS model display multi-stripped
configurations up to $t \sim 10^{5}$ mcs \cite{muka}.
It is then clear that the short time dynamics
must be studied using an order parameter which takes into account
multi-stripped configurations as that given by eq. (3).

Our results for the short time dynamical behavior
have been summarized in figure 4.
\begin{figure}
\centerline{\epsfig{file=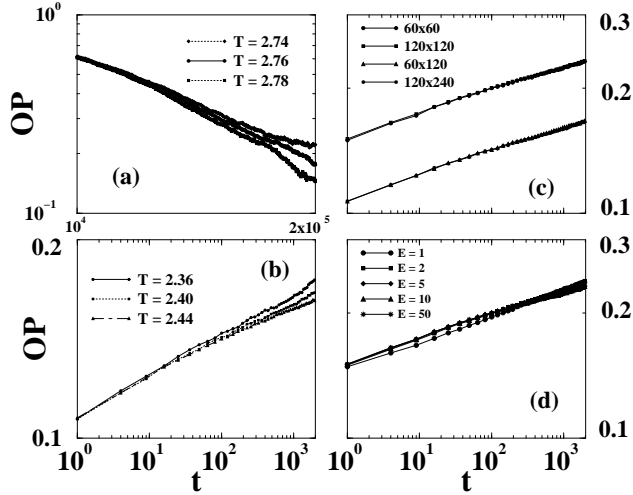, height=2.5in}}
\caption{Log-log plots of the OP versus $t$, as obtained by means of
the short time dynamics study. Averages are taking over
$10^3$ different runs using lattices of size 
$L_{x}/L_{y} = 2, L_{y} = 120$.
The following cases are shown: 
a) Starting from an ordered state, b) Starting from a fully 
disordered state, c) As in b) but for different lattice sizes, d)
as in b) but for different driving fields.}
\label{f.4}
\end{figure}
For the used density ($\rho_{o} = 0.16$) power laws have
been obtained for $T^{**} = 2.76$ (figure 4(a)) and    
$T^{*} = 2.40$ (figure 4(b)) starting from ordered and fully disordered
states, respectively. Also, figure 4(c) shows that the lower
bound given by the short time dynamics is independent
of the lattice size. Notice that the curves obtained for 
different aspect ratios are shifted but the power law behavior is
obtained at the same temperature. The same results have been obtained
for the upper bound, pointing out that the bounds 
drawn in the phase diagram (figure 3) are independent of the lattice
size and consequently also valid in the thermodynamic limit.
The transition points estimated using a finite lattice (figure 2)
satisfy  $T^{*} < T_{c} < T^{**}$ as would also do
the true transition points in the $L_{x}, L_{y} \rightarrow \infty$
limit. Also, the difference $\Delta T = T^{**} - T^{*}$ depends 
on the strength of the first-order transition while
$\Delta T \equiv 0$ at the second-order transition point 
for $\rho_{o} = 1/2$. 

Coming back to the phase diagram, it is found that for $\rho_o \ge 0.30$,
$T_c(E)$ steadily {\bf increases} with the strength of the 
field, reaching a saturation value at 
$T_c(E = \infty) \simeq 1.41 \, T_c(E=0)$ for $\rho_o = 1/2$, 
in excellent agreement with results for the KLS model \cite{ktl1,wan}. 
However, for lower densities (e.g. for $\rho_o < 0.1$, in
figure 3) $T_{c}(E)$ steadily {\bf decreases} when increasing the
magnitude of the field. So, $T_{c}(E)$ 
exhibits opposite trends depending on the density and 
consequently, it is 
expected that for some characteristic density $\rho_{o}^{M}$ 
($0.20 \geq \rho_{o}^{M} \geq 0.15$)
the critical temperature will be the same for all magnitudes 
of the driving field.
Therefore, the point $(\rho_{o}^{M}, T_c(E,\rho_{o}^{M}))$ is
a {\bf multicritical} point, in the sense that for these special values of
density and temperature this point is a critical point for all values
of the amplitude of the field. 
Due to the observed symmetry, 
$(1 - \rho_{o}^{M}, T_c(E,\rho_{o}^{M}))$ is also a multicritical point. 

Assuming that the critical curves have the
simplest form allowed by the symmetry of the system,
we propose the
following expression for the critical temperature: 
\begin{eqnarray}
T_c(E,\rho_{o}) &= T_c(\infty,1/2) - k_{\infty} f(E) 
(\frac{1}{2} \pm  \rho_{o}^{M})^{1/\beta} 
- k_{\infty} (1 - f(E)) (\frac{1}{2} \pm \rho_{o})^{1/\beta} ,\; E > 0,
\label{eq9}
\end{eqnarray}
where for $E \rightarrow \infty$, $k_{\infty}$ is the 
coefficient of the higher order term and $f(E)\rightarrow 0$, respectively.
Equation (\ref{eq9}) can be thought as the first approximation to
the phase coexistence curve, valid close to $\rho_{o} = 1/2$,
so that $\beta$ is the order parameter critical exponent
of the second-order transition.
In order to fit eq. (\ref{eq9}) to the data we will first summarize 
the symmetries present in our model. The model exhibits full translational 
and reflexion invariance as
the Ising model, but the rotational symmetry is broken due to the
anisotropy introduced by the field. If we consider short time scales,
the up-down symmetry is also broken by the field. However, a
renormalization group study will consider the system at a coarse-grained
level. Then, we expect that the up-down symmetry will be restored at
long time scales. Consequently, the present model displays the same
symmetries than the randomly driven lattice gas with  
$\beta = \frac{1}{3}$ \cite{grana1}. 
Taking this value for $\beta$,  the critical curve
for $E = \infty$ can be fitted using a single 
parameter, yielding $k_{\infty} = 15 \pm 3$. 
Assuming that $f(E) = \exp{(-E)}$, $\rho_{o}^{M}$ is the
only parameter left to be fitted, yielding $\rho_{o}^{M} = 0.160 \pm
0.005$ for $E = 1$ (see figure 4). Discrepancies between the fit
and the data for densities far from $\rho_o = 1/2$ are expected 
since the expansion given by eq. (\ref{eq9}) holds close to that
point only.
Notice that equation (\ref{eq9}) satisfies that 
$(\rho_{o}^{M}= 0.160 \pm 0.005, T_c(\rho_{o}^{M}) = 2.59 \pm 0.01 ))$ 
is a multicritical point. This value is in agreement with the
estimation performed using the short time dynamics study
that gives  $T^{**} = 2.76 > T_c(\rho_{o}^{M}) > T^{*} = 2.40$ 
(see figure 4). The existence of the multicritical point can
also be confirmed by means of a short time dynamics simulations. In fact, 
figure 4(d) shows that plots of $OP_{x}$ versus $t$ obtained for
different fields ($1 \leq E \leq \infty$) yield the
same lower bound for $T_{c}(\rho_{o}^{M})$ given by $T^{*} = 2.40$,
independently of the strength of the field. This behavior
is characteristic of the multicritical point, as  
observed in figure 3. It should be noticed that 
fits of the phase diagram assuming $\beta = 1/2$, as theoretically 
expected for the KLS model \cite{bea1}, are far from being satisfactory. 
Also, an excellent fit of the curve can be obtained assuming
$\beta = 1/4$ 
(yielding $(\rho_{o}^{M}= 0.168 \pm 0.005, 
T_c(\rho_{o}^{M}) = 2.57 \pm 0.01)$), but this value of the exponent
is not supported by the symmetry considerations above mentioned.  
  
For the sake of comparison we have included in the phase diagram the
critical temperature of the Ising model $T_{c}^{I}$ as well as the prediction
of eq. (\ref{eq9}) in the $E \rightarrow 0$ limit. The latter
is in excellent agreement with the lower bound estimate given by the
short time dynamics method for $E = 0.01$. Notice that these
estimations for the driven system are consistent with the 
location of the multicritical point that should also
hold for $E \rightarrow 0$. These results show that, for 
$\rho = 1/2$, there is a gap in the critical temperature between 
the case $E = 0$ (Ising model) and the limit $E \rightarrow 0$ of the
present model. Such a gap is expected to be even greater 
for $\rho \neq 1/2$ because in this case
the coexistence temperature of the Ising model 
is lower than $T_{c}^{I}$ while the coexistence temperature
of the driven diffusive system has a lower bound given
by the multicritical temperature. 
The existence of these temperature gaps dramatically reflects 
the anisotropy introduced by the driving field and the 
non-equilibrium nature of the studied model. A physical 
explanation of this observation remains as an open question.  

In summary, 
the phase diagram of a DDS in the presence of an oscillatory driving field
is determined for $E = 1$ and $E = \infty$. 
We give strong evidence on the existence of a 
multicritical point and a critical temperature gap
separating the cases $E = 0$ from $E \rightarrow 0$.
To our best knowledge,
these features have never been reported in the field
of DDS. 
\vskip 0.5 true cm
{\bf Acknowledgments}: This work was supported by CONICET, 
UNLP, ANPCyT  and Fundaci\'on Antorchas (Argentina). We 
acknowledge useful discussions with B. Schmittmann.

\end{document}